\documentclass[%
 reprint,
 amsmath,amssymb,
 aps,
prl,
]{revtex4-2}

\usepackage{graphicx}
\usepackage{dcolumn}
\usepackage{bm}
\usepackage{textgreek}

\begin{document}

\title{Classical and Quantum Phase Transitions in Multiscale Media:\\ Universality and Critical Exponents in the Fractional Ising Model}

\author{Joshua M. Lewis$^{1,2}$}
\author{Lincoln D. Carr$^{1,2,3}$}
\affiliation{$^1$Quantum Engineering Program, Colorado School of Mines, 1523 Illinois St., Golden, CO 80401, USA}
\affiliation{$^2$Department of Physics, Colorado School of Mines, 1523 Illinois St., Golden, CO 80401, USA}
\affiliation{$^3$Department of Applied Mathematics and Statistics, Colorado School of Mines, 1523 Illinois St., Golden, CO 80401, USA}

\date{\today}

\begin{abstract}
    Until now multiscale quantum problems have appeared to be out of reach at the many-body level relevant to strongly correlated materials and current quantum information devices. In fact, they can be modeled with $q$-th order fractional derivatives, as we demonstrate in this work, treating classical and quantum phase transitions in a fractional Ising model for $0 < q \leq 2$ ($q = 2$ is the usual Ising model). We show that fractional derivatives not only enable continuous tuning of critical exponents such as $\nu$, $\delta$, and $\eta$, but also define the Hausdorff dimension $H_D$ of the system tied geometrically to the anomalous dimension $\eta$.  We discover that for classical systems, $H_D$ is precisely equal to the fractional order $q$. In contrast, for quantum systems, $H_D$ deviates from this direct equivalence, scaling more gradually, driven by additional degrees of freedom introduced by quantum fluctuations. These results reveal how fractional derivatives fundamentally modify the fractal geometry of many-body interactions, directly influencing the universal symmetries of the system and overcoming traditional dimensional restrictions on phase transitions. Specifically, we find that for $q < 1$ in the classical regime and $q < 2$ in the quantum regime, fractional interactions allow phase transitions in one dimension. This work establishes fractional derivatives as a powerful tool for engineering critical behavior, offering new insights into the geometry of multiscale systems and opening avenues for exploring tunable quantum materials on NISQ devices.
\end{abstract}

\maketitle

The study of phase transitions has long been a cornerstone of condensed matter physics, providing profound insights into the behavior of matter at its most fundamental level. 
Most paradigmatic models for phase transition, such as the Ising model~\cite{Lenz1920, Ising1925, Stanley1971, Chandler1987, Adler1990}, assume nearest-neighbor interactions, aka tight binding.  However, extensive research has explored systems with varying interaction ranges, e.g. the power-law potentials inherent in many quantum simulators/emulators~\cite{jaschke2017critical, joshi2020quantum, gambetta2020long, gong2017entanglement}. Nearest-neighbor models represent only a special case of local interactions against a broader spectrum that includes long-range couplings, which can significantly influence the system's critical behavior~\cite{angelini2014relations, gonzalez2021finite}. This spectrum of interactions then has physical significance due to the ability to carefully engineer these models~\cite{grass2014trapped, stute2012tunable}, for instance, programmable long-range interactions in Rydberg atom chains utilized also in quantum computing~~\cite{secker2016controlled}.  In this Letter, we uncover a novel tunability in the critical exponents of classical and quantum phase transitions driven by medium- to long-range interactions in \emph{multiscale} quantum systems.  Such multiscale phenomena, described by \emph{fractional derivatives}, are quite common in large scale classical systems found in geophysics, soft condensed matter, and biophysics~\cite{zhang2017review, chen2008intuitive, nonnenmacher2000applications}; here we describe their effects on correlated quantum materials.  Fractional effects not only overcome dimensional restrictions on the existence of critical points but also reveal unique phenomena in the quantum phase transition regime, fundamentally modifying the underlying fractal dimension of many-body interactions, and thereby directly influencing the universal symmetries of the model and opening a new axis for exploring critical phenomena in quantum materials.


Fundamental to the study of phase transitions, \emph{scale invariance} at the critical point and its connection to system geometry is pivotal to characterizing the universality class of the model, regardless of whether it  features short- or long-range interactions. The \emph{Hausdorff dimension} $H_D$ provides a quantitative measure of how a given set fills space, capturing the fractal nature of structures that arise at criticality, providing a rich geometric interpretation for the scale-invariant critical fluctuations that arise at the phase transition. In this context, $H_D$ affects how physical quantities like correlation functions, transport properties, and susceptibility behave near critical points. It influences the scaling laws and critical exponents that characterize the system's response to external perturbations and determines the geometry of critical clusters and domains.  The fractional derivative has emerged as a powerful tool for modeling nonlocality and long-range (but non-power-law) interactions inherent in multiscale complex systems. This derivative intrinsically incorporates a nonlocal kernel, capturing scale-invariant behaviors that are directly linked to the fractal geometry characterized by $H_D$.  The fractional derivative has a number of formulations; perhaps the most familiar to physicists is the Reisz formulation, defining $d^q f(x)/d |x|^q$ via its Fourier transform, $-|k|^q \tilde{f}(k)$.






In quantum mechanics, the introduction of fractional derivatives into the usual zero point energy or dispersive term leads to new insights in Laskin's reformulation~\cite{laskin2000fractional, laskin2000fractionalQ, laskin2002fractional, stickler2013potential, hasan2018tunneling, zhang2015propagation}. Laskin's work generalizes single-particle quantum mechanics by extending the Feynman path integral to include Lévy flights rather than those typical of Brownian trajectories. Whereas Brownian trajectories trace out paths that form a fractal object with $H_D=2$ in two-dimensional space, Lévy flights generalize this concept by allowing for arbitrary $H_D$. Lévy flights arise from stochastic processes with a jump length probability density function that behaves asymptotically as $\mathcal{P}_q(x) \propto 1/|x|^{q + 1}$, where $q$ is the Lévy stability index.
%
The Lévy flights giving rise to fractional derivatives are not merely mathematical curiosities but are widely used to describe anomalous scaling in complex systems, including the dynamical correlations in 1D Hamiltonian systems~\cite{mendl2015current, van2012exact, dhar2013exact, kundu2019fractional}, particle motion in turbulent flows~\cite{solomon1993observation}, non-Newtonian viscosity~\cite{pandey2016linking}, animal movement patterns~\cite{brockmann2006scaling, benhamou2007many, murakami2019levy}, neuron signaling~\cite{liu2021levy}, and financial market dynamics~\cite{yarahmadi20222d}. Single-particle quantum versions of these scalings have also found applications in optical systems~\cite{iomin2021fractional, zeng2019one, xin2021propagation, he2021propagation, zhang2015propagation}, with recent experimental realizations using a Lévy waveguide~\cite{liu2023experimental}.    In this Letter we treat the many-body generalization of these notions, with rather remarkable consequences, in a \emph{L\'evy crystal}~\cite{stickler2013potential, zhang2015propagation}.

\emph{Lévy Crystal} --- The 1D space-fractional Landau-Ginzburg functional in free space is given by, 
\begin{equation}
    H = \int dx \left[ -\frac{1}{2} K\, \phi(x) \partial_x^q\phi(x) + h\, \phi(x) \right]\,,
\end{equation}
%
in which $q$ is the fractional order of the derivative, $K$ characterizes the strength of the order-driving fractional derivative, and $h$ represents the external biasing field. For simplicity, we employ natural units in the dispersive term and focus on the effects of the fractional derivative itself, recognizing that different constants may be required for varying fractional orders to maintain consistent units. However, these constants are not universal and thus do not influence the scaling behavior of the model. Typically, the order of the derivative $q$ is restricted to the domain $q \in (1,2]$. However, due to the existence of biological systems with Lévy indices in the range $q \in (0,1)$, and recent optical experiments using Lévy waveguides to realize Lévy flights with $q \in (0,1)$~\cite{liu2023experimental}, we explore this extended parameter regime for Lévy indices in the broader domain $q \in (0,2]$.

\begin{figure}[t]
    \centering
    \includegraphics[width=\columnwidth]{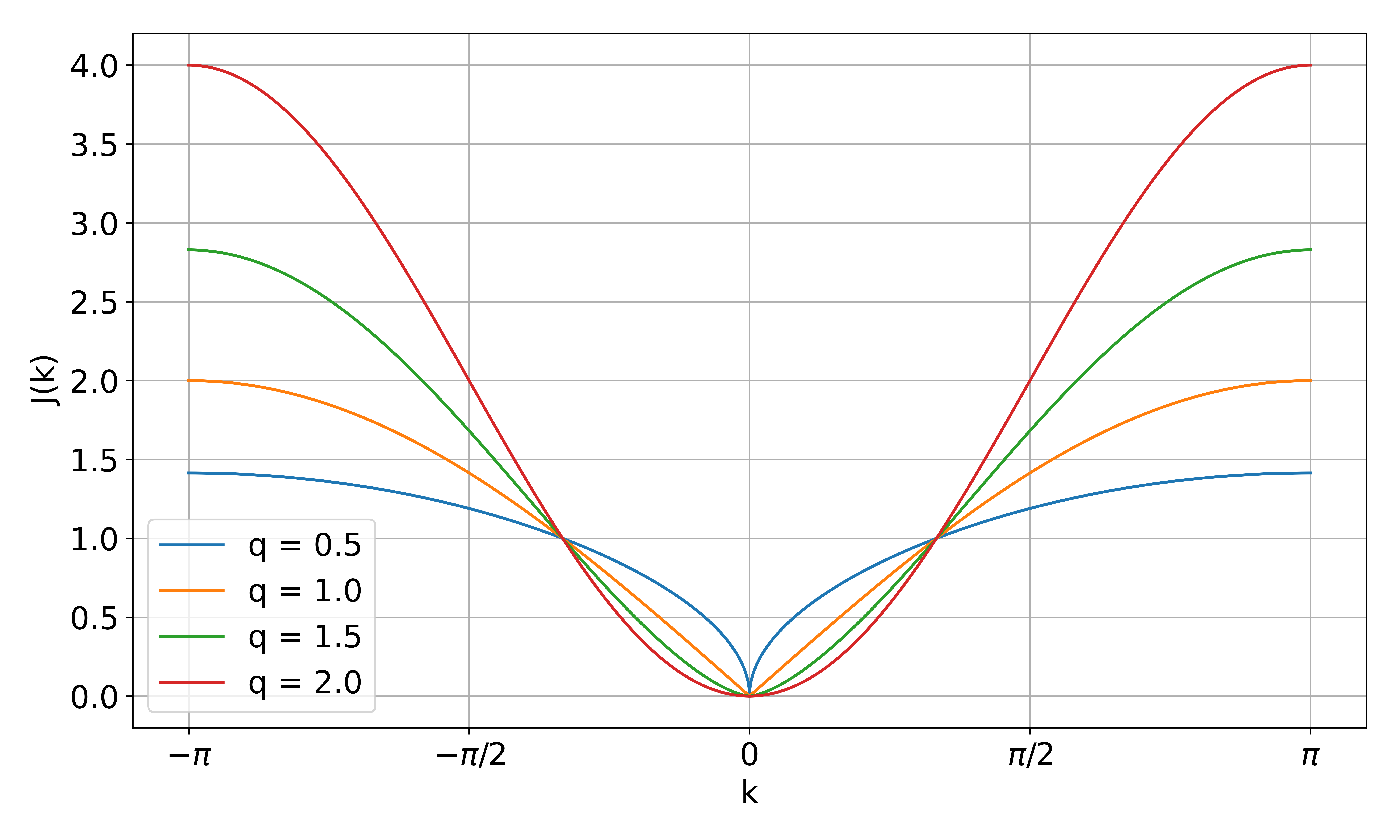}
    \caption{Momentum-space representation of fractional interactions on a periodic lattice given by Eq.~(\ref{eq:momentumInteractions}), corresponding to a first Brillouin zone spanning $-\pi$ to $\pi$. As $q$ decreases, the interactions become increasingly more weighted near $k = 0$, indicating a broadening of the effective interaction range in real space. This enhancement at low $q$ reflects the dominance of long-wavelength contributions, fundamentally altering the system's critical behavior by increasing the coupling strength of modes with small momentum.}

    \label{fig:interactionsAcrossOrderMomentum}
\end{figure}

The Riesz fractional derivative operator may be written as a finite difference to second order as $\Delta_a^q$ in which $a$ acts as a finite difference spacing using the operator proposed by Ortigueira~\cite{ortigueira2006riesz}:
\begin{eqnarray}
    \Delta_a^q\psi(x) &\equiv& \sum_{j=-\infty}^{\infty} (-1)^j\binom{q}{\frac{q}{2} + j}\psi(x - ja)\,,\\
    \label{secondOrderFD}
    \partial_x^q\psi(x) &\approx& - \frac{\Delta_{a}^{q}}{a^q}\psi(x) + \mathcal{O}(a^2)\,.
\end{eqnarray}
A similar expression holds for the Fourier representation, 
with the Fourier wave-number on an infinite domain $k$ mapping to a periodic lattice to $\frac{2}{a} \sin\left(\frac{k a}{2}\right)$, with $a$ as the lattice constant. Performing this substitution with a unit lattice constant, the interactions in the momentum domain are given by 
\begin{equation}
    \tilde{J}(k) = \left|2\sin\left(\frac{k}{2}\right)\right|^q \,,
    \label{eq:momentumInteractions}
\end{equation}
see Fig.~(\ref{fig:interactionsAcrossOrderMomentum}). Importantly, the continuous form $ |k|^q $ is recovered in the small $k$ limit, the long-wavelength regime, which governs critical behavior near phase transitions.  For $q = 2$, Eq.~(\ref{secondOrderFD}) simplifies to the standard second-order derivative, which in the context of many-body interactions corresponds to a model with nearest-neighbor terms, aka tight binding. 

By applying the general discretization of Eq.~(\ref{secondOrderFD}) to a lattice, one obtains the Lévy crystal.  In the case of the Ising model, the spin-spin coupling between spins $i$, $j$ takes the form
\begin{equation}
    J(r) = (-1)^{r + 1}\binom{q}{\frac{q}{2} + r}\,,
\label{eq:coupling}
\end{equation}
with $r=|i-j|$.
\begin{figure}[t]
    \centering
    \includegraphics[width=\columnwidth]{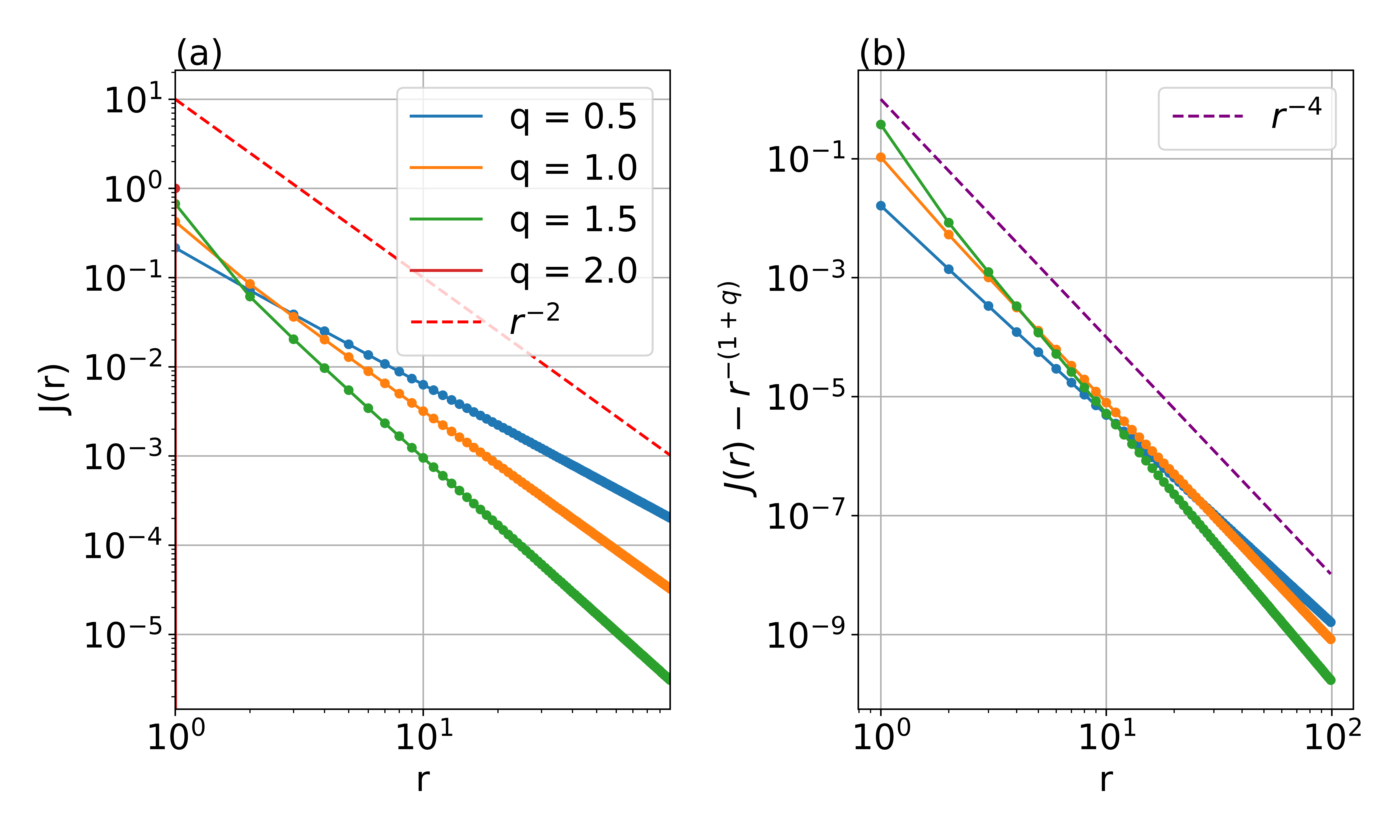}
    \caption{(a) Fractional interactions as a function of distance for varying fractional orders q. The interactions exhibit a clear power-law decay, with the leading exponent increasing monotonically with q. For comparison, a reference curve of $r^{-2}$ is included to illustrate the asymptotic behavior relative to the classical interaction decay. (b) Residual interactions are obtained by subtracting the leading asymptotic term, revealing subleading contributions that follow a power-law decay. An $r^{-4}$ reference curve is included for comparison.}
    \label{fig:interactionsAcrossOrder}
\end{figure}
Asymptotically, Eq.~(\ref{eq:coupling}) approaches a power law of form 
\begin{equation}
    J(r) \sim r^{-(1+q)} + r^{-(3+q)} \,,
    \label{eq:assymtoticCoupling}
\end{equation}
see Fig.~(\ref{fig:interactionsAcrossOrder}). However, at short to medium range, Eq.~(\ref{eq:coupling}) contains additional components not observed in Eq.~\ref{eq:assymtoticCoupling}.  In fact, for fractional orders $q>2$ the structure of the interactions changes fundamentally at short to medium range, introducing a nearest-neighbor ferromagnetic interaction alongside an additional \emph{antiferromagnetic} power law, which in turn breaks up the ferromagnetic order at large scales. 

The Hamiltonian for the multiscale 1D quantum fractional transverse-field Ising model is defined as
\begin{equation}
    \hat{H} = -J_0\sum_{i<j}J(|j - i|)\hat{\sigma}_i^z \hat{\sigma}_j^z + g\sum_{j}\hat{\sigma}_j^x + h\sum_{j}\hat{\sigma}_j^z \,.
\end{equation}
where $J_0$ is a constant for the overall strength of the interactions, $J(|j-i|)$ is the fractional interaction distribution, and $g$ and $h$ represent the strength of the transverse and longitudinal biasing fields, respectively. 

\emph{Finite Size Scaling} --- To measure the anomalous scaling induced by the fractional derivative, we examine the six critical exponents that define the unique universality class of the model. They are determined by measuring five key quantities across different system sizes and temperatures within a narrow scaling window around the critical point: the magnetization $M$, susceptibility $\chi$, specific heat $C$, Binder's cumulant $U$, and the connected spin-spin correlation function $G(r)$, 
\begin{equation}
    \begin{aligned}
        M & = \langle m \rangle \\
        \chi & = L(\langle m^2 \rangle - \langle m \rangle^2) \\
        C & = \frac{1}{k_B T^2}(\langle E^2 \rangle - \langle E \rangle^2) \\
        U & = 1 - \frac{\langle m^4 \rangle}{3\langle m^2 \rangle} \\
        G(r) & = \langle S_i S_{i+r} \rangle - \langle S_i \rangle \langle S_{i+r} \rangle \,.
    \end{aligned}
    \label{eq:observables}
\end{equation}

According to well-established finite size scaling theory~\cite{privman1990finite} these quantities behave as a function of a changing system size $L$, reduced temperature $\epsilon=T/T_c$, with $T_c$ as the critical temperature, and reduced biasing field $h$,
\begin{equation}
    \begin{aligned}
        M(L, \epsilon) & = L^{-\beta/\nu}\tilde{M}(L^{1/\nu}\epsilon), \\
        M_h(L, h) & = L^{-\beta/\nu} \tilde{M}_h(h L^{\delta/\nu}), \\
        \chi(L, \epsilon) & = L^{\gamma/\nu}\tilde{\chi}(L^{1/\nu}\epsilon), \\
        C(L, \epsilon) & = L^{\alpha/\nu}\tilde{C}(L^{1/\nu}\epsilon), \\
        U(L, \epsilon) & = \tilde{U}(L^{1/\nu}\epsilon), \\
        G(r, L, \epsilon) & = L^{-(d-2+\eta)}\tilde{G}\left(\frac{r}{L}, L^{1/\nu}\epsilon\right)\,,
    \end{aligned}
\end{equation}
where the tilde indicates the universal scaling function of the observable in a small scaling window around the critical temperature.  With these measurements, we extract the six critical exponents associated with the correlation length $(\nu)$, critical isotherm $(\delta)$, magnetization $(\beta)$, susceptibility $(\gamma)$, specific heat $(\alpha)$, and anomalous dimension $(\eta)$.  

The upper critical dimension of the 1D fractional Ising model can be defined analogously to power-law models as $d_u = 2q$. For fractional orders below 0.5, we must adjust our scaling hypothesis to account for scaling above the upper critical dimension. In this regime, the scaling hypothesis proposed by Flores et al.~\cite{flores2015finite} introduces a relaxation of the assumption that the correlation length is bounded by the system size $L$. This leads to a new scaling hypothesis characterized by an additional critical exponent $\kappa$, which rescales the exponents to account for interactions that behave as if they exist in a lower dimension than the underlying model. For an arbitrary observable $P$, whose original scaling theory follows the power law $P(L, \epsilon) \sim L^{p/\nu}$, the modified scaling is given by
\begin{equation}
    P(L, \epsilon) \sim L^{p \kappa/\nu} \,.
\end{equation}
When below the upper critical dimension, the exponent $\kappa$ is then defined as Eq.~(\ref{eq:kappaDefinition}), representing the effective cover of the space through interactions. 
\begin{equation}
    \kappa =
    \begin{cases} 
    d/d_{u} & \text{if } d_{u} < d, \\
    1 & \text{otherwise}.
    \end{cases}
    \label{eq:kappaDefinition}
\end{equation}
We utilize this definition when performing a finite-size scaling analysis where the model dimension is above the upper critical dimension, $d_{u}$.

\begin{figure}[!htbp]
    \centering
    \includegraphics[width=\columnwidth]{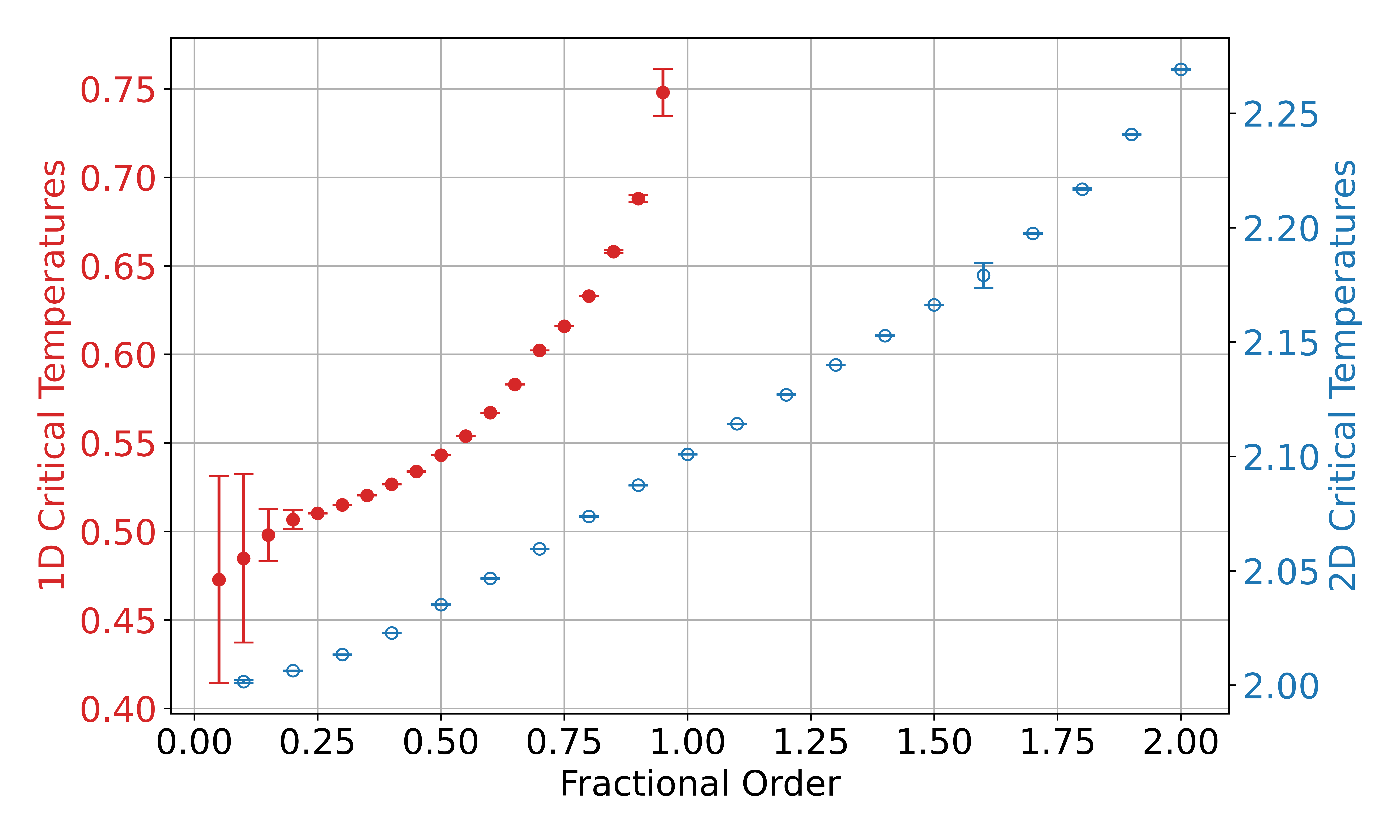}
    \caption{ Extrapolated thermodynamic critical temperatures for the 1D quantum and 2D classical Ising models. As $q$ increases, energy penalties at higher wavenumbers intensify, escalating disorder instability, while penalties for $|k| < 1$ (excluding $k=0$) decrease, enhancing stability for smaller variations. The increase in the thermodynamic critical point with $q$ reflects the scaling theory where the critical temperature scales with an exponent $1/\nu$, making extrapolation increasingly challenging as $\nu$ diverges and larger uncertainties emerge, particularly for the 1D case as $q \to 0$.}
    \label{fig:thermodynamicCriticalPoint}
\end{figure}

Each observable defined in Eq.~(\ref{eq:observables}) exhibits a peak at a pseudo-critical temperature, corresponding to the dominance of critical fluctuations associated with the phase transition for a given system size. By systematically tracking these pseudo-critical temperatures across varying system sizes, we extrapolate their behavior to determine the thermodynamic critical temperature in the limit of an infinite system size. The results for both the classical and quantum phase transitions are presented in Fig.~\ref{fig:thermodynamicCriticalPoint}.

\begin{figure*}[!htbp]
    \centering
    \includegraphics[width=\textwidth]{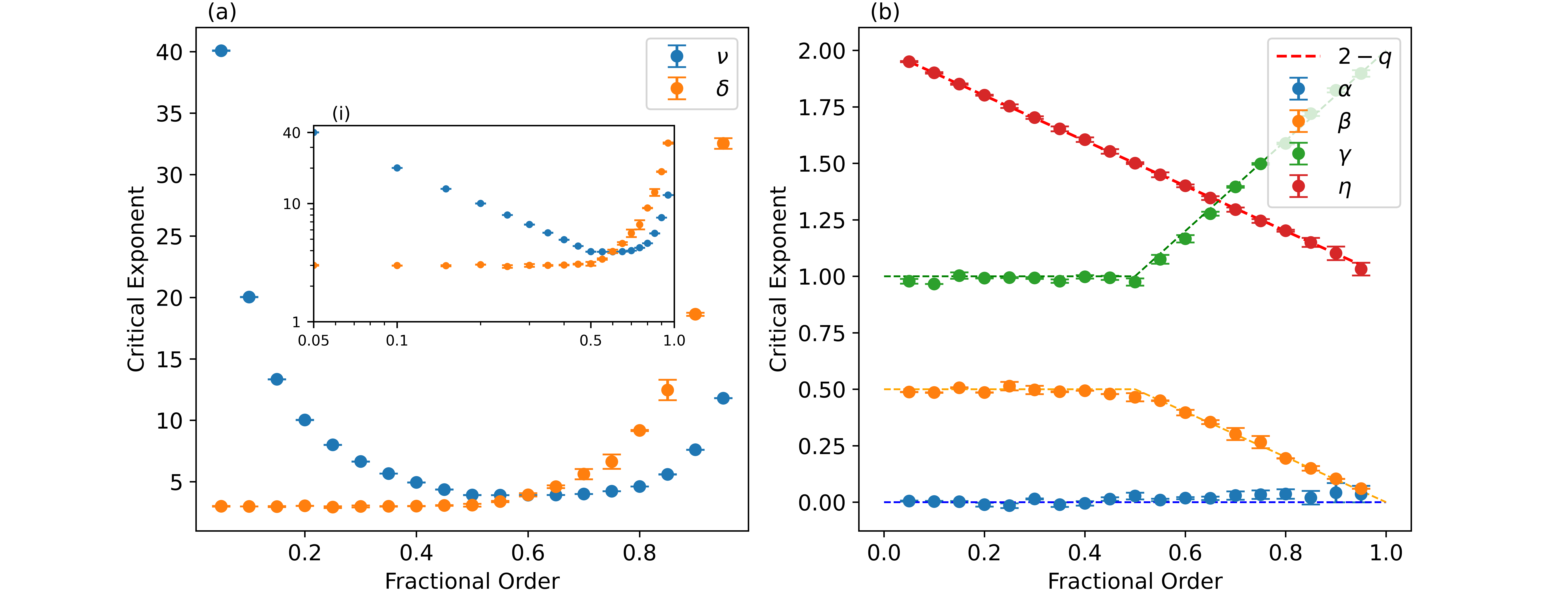}
    \caption{ Critical exponents $\nu$ and $\delta$ as functions of the fractional order $q$ for the classical phase transition, demonstrating continuous variation over the range $0 < q \leq 1$. The inset (i) highlights the behavior at $q = 0$, where $\nu$ diverges following a power-law scaling of $1/q$. (b) Critical exponents $\alpha$, $\beta$, $\gamma$, and $\eta$, including a functional form for $\eta$, illustrating its dependence on $q$. $\alpha$ remains invariant at $0$ across all $q$, while $\beta$ exhibits monotonic decay toward $0$ as $q$ increases. Conversely, $\gamma$ rises steadily, reaching a value of $2$ at $q = 1$.  For $q < 0.5$, $\gamma$ and $\beta$ retain their mean-field values, whereas beyond this threshold, all exponents besides the specific heat critical exponent $\alpha$ show clear variation. }

    \label{fig:classicalCriticalExponents}
\end{figure*}

\emph{Classical Phase Transition} --- In Fig.~\ref{fig:classicalCriticalExponents}, we present the critical exponents obtained from the classical phase transition, spanning fractional orders from $q=0$ to 1. We observe that critical exponents vary continuously with $q$, indicating that the fractional order serves as a marginal parameter---one that alters the universality class of the system as it changes. Of particular interest is the simple form of the anomalous dimension $\eta$, which connects this behavior to the geometric interpretation of the critical exponent.

Hove et al.~\cite{hove2000hausdorff} demonstrated that the anomalous dimension critical exponent $\eta$ and the Hausdorff dimension $H_D$ share a dual relationship,
\begin{equation}
    \eta + H_D = 2 \,.
\end{equation}
We now establish our scaling hypothesis developed from the data found in Fig.~\ref{fig:classicalCriticalExponents} that $\eta(q) = 2 - q$, implying,
\begin{equation}
    H_D = q \,.
    \label{HausdorffFractionalConnection}
\end{equation}
Based on our calculations of the anomalous dimension critical exponents, we conclude, with a $\chi^2$ statistic corresponding to a confidence level of 3.8 sigma (based on bootstrapping to calculate a variance in $\eta$), that the fractional derivative in the 1D Ising phase transition serves to precisely tune $H_D$ of the underlying multiscale structure. The observed modifications in other critical exponents follow directly from this new geometric framework.

\begin{figure*}[!htbp]
    \centering
    \includegraphics[width=\textwidth]{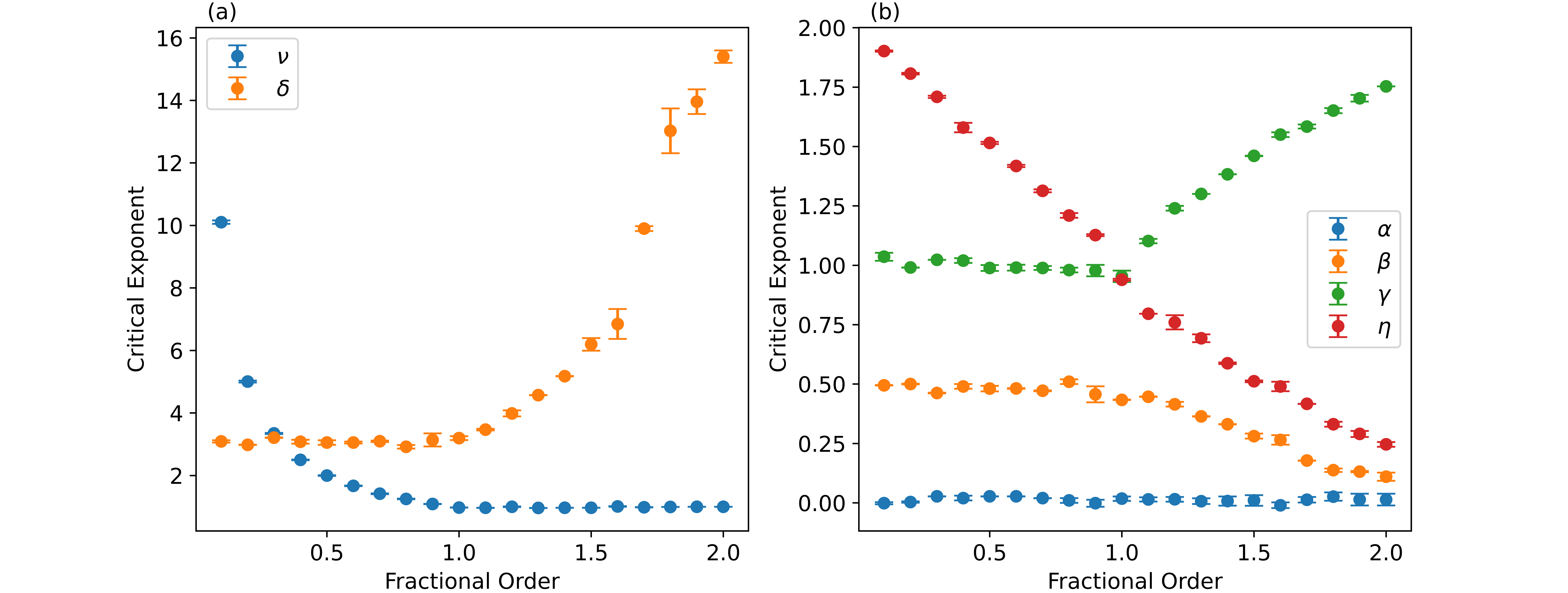}
    \caption{ (a) Critical exponent $\nu$ as a function of fractional order $q$ for the quantum phase transition. Similar to the the classical phase transition Kosterlitz-Thouless (KT) behavior, where $\nu$ diverges is observed at $q = 0$, though is absent at our upper bounding fractional order. (b) Critical exponents $\alpha$, $\beta$, $\gamma$, and $\eta$ as functions of $q$, displaying trends similar to those observed in the classical phase transition. However, the rate of change of the critical exponents with respect to $q$ is notably slower in the quantum case with a covariance of approximately 0.75 in comparison to the classical phase transition.}

    \label{fig:quantumCriticalExponents}
\end{figure*}

\emph{Quantum Phase Transition} --- Turning to the quantum case, we utilize the correspondence between the critical exponents of a $d$-dimensional quantum phase transition and a $(d+1)$-dimensional classical one~\cite{vojta2003quantum} and present these in Fig.~\ref{fig:quantumCriticalExponents}. While the classical phase transition occurs for fractional orders between 0 and 1, the additional degrees of freedom introduced in the quantum phase transition result in a a wider range of allowed fractional orders ranging from 0 to 2.

For the quantum phase transition, the relation in Eq.~(\ref{HausdorffFractionalConnection}) requires a corrective factor to account for the shifting balance between entropic disorder effects and the ordering induced by the fractional derivative in the higher-dimensional model. We find that in this quantum model, the covariance for $H_D$ varying with respect to the fractional order is approximately 0.75, compared to the unit covariance observed for the classical phase transition. 

Finally, we observe that in the regime where the fractional order $q$ is less than $d/2$, the susceptibility ($\gamma$) and magnetization ($\beta$) critical exponents freeze to their mean-field values, while the others continue to vary continuously. This discrepancy arises because these two exponents are more directly tied to the dominant spatial scale of fluctuations in the system. For $q<d/2$, the effective range of interactions is short enough that fluctuations are governed primarily by local interactions, which are insensitive to changes in $q$ within this regime. In contrast, other critical exponents, such as $\nu$ and $\eta$, are more sensitive to the global connectivity of interactions and the fractal geometry introduced by the fractional derivative, allowing them to vary continuously even when $q$ is below $d/2$. This highlights the fractional derivative’s dual role: as a weaker relevant parameter for local critical exponents below $d/2$, and as a stronger marginal parameter for global exponents above this threshold.

In conclusion, our study of the fractional Ising model demonstrates that fractional derivatives provide precise control over critical behavior in both classical and quantum phase transitions. The fractional order $q$ acts as a continuous marginal parameter, tuning critical exponents, reshaping system geometry through the anomalous dimension, and modifying effective dimensionality by its connection to the critical exponent $\eta$. Extending this framework to quantum phase transitions for $q \leq 2$, we highlight how the additional degrees of freedom affect scaling behavior and critical phenomena. These findings establish a direct link between fractional derivatives, the Hausdorff dimension, and the emergence of order, offering insights into multiscale structures and potential applications to engineered quantum systems and materials.

\emph{Acknowledgments} -- This work was performed in part while at the Kavli Institute for Theoretical Physics at the University of California, Santa Barbara;  and supported in part by the National Science Foundation under grants PHY-2210566, DGE-2125899, and DMR-2002980.

\bibliography{sources}

\begin{thebibliography}{44}%
\makeatletter
\providecommand \@ifxundefined [1]{%
 \@ifx{#1\undefined}
}%
\providecommand \@ifnum [1]{%
 \ifnum #1\expandafter \@firstoftwo
 \else \expandafter \@secondoftwo
 \fi
}%
\providecommand \@ifx [1]{%
 \ifx #1\expandafter \@firstoftwo
 \else \expandafter \@secondoftwo
 \fi
}%
\providecommand \natexlab [1]{#1}%
\providecommand \enquote  [1]{``#1''}%
\providecommand \bibnamefont  [1]{#1}%
\providecommand \bibfnamefont [1]{#1}%
\providecommand \citenamefont [1]{#1}%
\providecommand \href@noop [0]{\@secondoftwo}%
\providecommand \href [0]{\begingroup \@sanitize@url \@href}%
\providecommand \@href[1]{\@@startlink{#1}\@@href}%
\providecommand \@@href[1]{\endgroup#1\@@endlink}%
\providecommand \@sanitize@url [0]{\catcode `\\12\catcode `\$12\catcode `\&12\catcode `\#12\catcode `\^12\catcode `\_12\catcode `\%12\relax}%
\providecommand \@@startlink[1]{}%
\providecommand \@@endlink[0]{}%
\providecommand \url  [0]{\begingroup\@sanitize@url \@url }%
\providecommand \@url [1]{\endgroup\@href {#1}{\urlprefix }}%
\providecommand \urlprefix  [0]{URL }%
\providecommand \Eprint [0]{\href }%
\providecommand \doibase [0]{https://doi.org/}%
\providecommand \selectlanguage [0]{\@gobble}%
\providecommand \bibinfo  [0]{\@secondoftwo}%
\providecommand \bibfield  [0]{\@secondoftwo}%
\providecommand \translation [1]{[#1]}%
\providecommand \BibitemOpen [0]{}%
\providecommand \bibitemStop [0]{}%
\providecommand \bibitemNoStop [0]{.\EOS\space}%
\providecommand \EOS [0]{\spacefactor3000\relax}%
\providecommand \BibitemShut  [1]{\csname bibitem#1\endcsname}%
\let\auto@bib@innerbib\@empty
\bibitem [{\citenamefont {Lenz}(1920)}]{Lenz1920}%
  \BibitemOpen
  \bibfield  {author} {\bibinfo {author} {\bibfnamefont {W.}~\bibnamefont {Lenz}},\ }\bibfield  {title} {\bibinfo {title} {Beiträge zum verständnis der magnetischen eigenschaften in festen körpern},\ }\href@noop {} {\bibfield  {journal} {\bibinfo  {journal} {Physikalische Zeitschrift}\ }\textbf {\bibinfo {volume} {21}},\ \bibinfo {pages} {613} (\bibinfo {year} {1920})}\BibitemShut {NoStop}%
\bibitem [{\citenamefont {Ising}(1925)}]{Ising1925}%
  \BibitemOpen
  \bibfield  {author} {\bibinfo {author} {\bibfnamefont {E.}~\bibnamefont {Ising}},\ }\bibfield  {title} {\bibinfo {title} {Beitrag zur theorie des ferromagnetismus},\ }\href {https://doi.org/10.1007/BF02980577} {\bibfield  {journal} {\bibinfo  {journal} {Zeitschrift für Physik}\ }\textbf {\bibinfo {volume} {31}},\ \bibinfo {pages} {253} (\bibinfo {year} {1925})}\BibitemShut {NoStop}%
\bibitem [{\citenamefont {Stanley}(1971)}]{Stanley1971}%
  \BibitemOpen
  \bibfield  {author} {\bibinfo {author} {\bibfnamefont {H.~E.}\ \bibnamefont {Stanley}},\ }\href@noop {} {\emph {\bibinfo {title} {Introduction to Phase Transitions and Critical Phenomena}}}\ (\bibinfo  {publisher} {Oxford University Press},\ \bibinfo {year} {1971})\BibitemShut {NoStop}%
\bibitem [{\citenamefont {Chandler}(1987)}]{Chandler1987}%
  \BibitemOpen
  \bibfield  {author} {\bibinfo {author} {\bibfnamefont {D.}~\bibnamefont {Chandler}},\ }\href@noop {} {\emph {\bibinfo {title} {Introduction to Modern Statistical Mechanics}}}\ (\bibinfo  {publisher} {Oxford University Press},\ \bibinfo {year} {1987})\BibitemShut {NoStop}%
\bibitem [{\citenamefont {Adler}(1990)}]{Adler1990}%
  \BibitemOpen
  \bibfield  {author} {\bibinfo {author} {\bibfnamefont {R.~J.}\ \bibnamefont {Adler}},\ }\href {https://doi.org/10.1214/lnms/1215465621} {\emph {\bibinfo {title} {An Introduction to Continuity, Extrema, and Related Topics for General Gaussian Processes}}},\ \bibinfo {series} {Lecture Notes--Monograph Series}, Vol.~\bibinfo {volume} {12}\ (\bibinfo  {publisher} {Institute of Mathematical Statistics},\ \bibinfo {year} {1990})\BibitemShut {NoStop}%
\bibitem [{\citenamefont {Jaschke}\ \emph {et~al.}(2017)\citenamefont {Jaschke}, \citenamefont {Maeda}, \citenamefont {Whalen}, \citenamefont {Wall},\ and\ \citenamefont {Carr}}]{jaschke2017critical}%
  \BibitemOpen
  \bibfield  {author} {\bibinfo {author} {\bibfnamefont {D.}~\bibnamefont {Jaschke}}, \bibinfo {author} {\bibfnamefont {K.}~\bibnamefont {Maeda}}, \bibinfo {author} {\bibfnamefont {J.~D.}\ \bibnamefont {Whalen}}, \bibinfo {author} {\bibfnamefont {M.~L.}\ \bibnamefont {Wall}},\ and\ \bibinfo {author} {\bibfnamefont {L.~D.}\ \bibnamefont {Carr}},\ }\bibfield  {title} {\bibinfo {title} {Critical phenomena and kibble--zurek scaling in the long-range quantum ising chain},\ }\href@noop {} {\bibfield  {journal} {\bibinfo  {journal} {New Journal of Physics}\ }\textbf {\bibinfo {volume} {19}},\ \bibinfo {pages} {033032} (\bibinfo {year} {2017})}\BibitemShut {NoStop}%
\bibitem [{\citenamefont {Joshi}\ \emph {et~al.}(2020)\citenamefont {Joshi}, \citenamefont {Elben}, \citenamefont {Vermersch}, \citenamefont {Brydges}, \citenamefont {Maier}, \citenamefont {Zoller}, \citenamefont {Blatt},\ and\ \citenamefont {Roos}}]{joshi2020quantum}%
  \BibitemOpen
  \bibfield  {author} {\bibinfo {author} {\bibfnamefont {M.~K.}\ \bibnamefont {Joshi}}, \bibinfo {author} {\bibfnamefont {A.}~\bibnamefont {Elben}}, \bibinfo {author} {\bibfnamefont {B.}~\bibnamefont {Vermersch}}, \bibinfo {author} {\bibfnamefont {T.}~\bibnamefont {Brydges}}, \bibinfo {author} {\bibfnamefont {C.}~\bibnamefont {Maier}}, \bibinfo {author} {\bibfnamefont {P.}~\bibnamefont {Zoller}}, \bibinfo {author} {\bibfnamefont {R.}~\bibnamefont {Blatt}},\ and\ \bibinfo {author} {\bibfnamefont {C.~F.}\ \bibnamefont {Roos}},\ }\bibfield  {title} {\bibinfo {title} {Quantum information scrambling in a trapped-ion quantum simulator with tunable range interactions},\ }\href@noop {} {\bibfield  {journal} {\bibinfo  {journal} {Physical Review Letters}\ }\textbf {\bibinfo {volume} {124}},\ \bibinfo {pages} {240505} (\bibinfo {year} {2020})}\BibitemShut {NoStop}%
\bibitem [{\citenamefont {Gambetta}\ \emph {et~al.}(2020)\citenamefont {Gambetta}, \citenamefont {Zhang}, \citenamefont {Hennrich}, \citenamefont {Lesanovsky},\ and\ \citenamefont {Li}}]{gambetta2020long}%
  \BibitemOpen
  \bibfield  {author} {\bibinfo {author} {\bibfnamefont {F.~M.}\ \bibnamefont {Gambetta}}, \bibinfo {author} {\bibfnamefont {C.}~\bibnamefont {Zhang}}, \bibinfo {author} {\bibfnamefont {M.}~\bibnamefont {Hennrich}}, \bibinfo {author} {\bibfnamefont {I.}~\bibnamefont {Lesanovsky}},\ and\ \bibinfo {author} {\bibfnamefont {W.}~\bibnamefont {Li}},\ }\bibfield  {title} {\bibinfo {title} {Long-range multibody interactions and three-body antiblockade in a trapped rydberg ion chain},\ }\href@noop {} {\bibfield  {journal} {\bibinfo  {journal} {Physical Review Letters}\ }\textbf {\bibinfo {volume} {125}},\ \bibinfo {pages} {133602} (\bibinfo {year} {2020})}\BibitemShut {NoStop}%
\bibitem [{\citenamefont {Gong}\ \emph {et~al.}(2017)\citenamefont {Gong}, \citenamefont {Foss-Feig}, \citenamefont {Brand{\~a}o},\ and\ \citenamefont {Gorshkov}}]{gong2017entanglement}%
  \BibitemOpen
  \bibfield  {author} {\bibinfo {author} {\bibfnamefont {Z.-X.}\ \bibnamefont {Gong}}, \bibinfo {author} {\bibfnamefont {M.}~\bibnamefont {Foss-Feig}}, \bibinfo {author} {\bibfnamefont {F.~G.}\ \bibnamefont {Brand{\~a}o}},\ and\ \bibinfo {author} {\bibfnamefont {A.~V.}\ \bibnamefont {Gorshkov}},\ }\bibfield  {title} {\bibinfo {title} {Entanglement area laws for long-range interacting systems},\ }\href@noop {} {\bibfield  {journal} {\bibinfo  {journal} {Physical review letters}\ }\textbf {\bibinfo {volume} {119}},\ \bibinfo {pages} {050501} (\bibinfo {year} {2017})}\BibitemShut {NoStop}%
\bibitem [{\citenamefont {Angelini}\ \emph {et~al.}(2014)\citenamefont {Angelini}, \citenamefont {Parisi},\ and\ \citenamefont {Ricci-Tersenghi}}]{angelini2014relations}%
  \BibitemOpen
  \bibfield  {author} {\bibinfo {author} {\bibfnamefont {M.~C.}\ \bibnamefont {Angelini}}, \bibinfo {author} {\bibfnamefont {G.}~\bibnamefont {Parisi}},\ and\ \bibinfo {author} {\bibfnamefont {F.}~\bibnamefont {Ricci-Tersenghi}},\ }\bibfield  {title} {\bibinfo {title} {Relations between short-range and long-range ising models},\ }\href@noop {} {\bibfield  {journal} {\bibinfo  {journal} {Physical Review E}\ }\textbf {\bibinfo {volume} {89}},\ \bibinfo {pages} {062120} (\bibinfo {year} {2014})}\BibitemShut {NoStop}%
\bibitem [{\citenamefont {Gonzalez~Lazo}\ \emph {et~al.}(2021)\citenamefont {Gonzalez~Lazo}, \citenamefont {Heyl}, \citenamefont {Dalmonte},\ and\ \citenamefont {Angelone}}]{gonzalez2021finite}%
  \BibitemOpen
  \bibfield  {author} {\bibinfo {author} {\bibfnamefont {E.}~\bibnamefont {Gonzalez~Lazo}}, \bibinfo {author} {\bibfnamefont {M.}~\bibnamefont {Heyl}}, \bibinfo {author} {\bibfnamefont {M.}~\bibnamefont {Dalmonte}},\ and\ \bibinfo {author} {\bibfnamefont {A.}~\bibnamefont {Angelone}},\ }\bibfield  {title} {\bibinfo {title} {Finite-temperature critical behavior of long-range quantum ising models},\ }\href@noop {} {\bibfield  {journal} {\bibinfo  {journal} {SciPost Physics}\ }\textbf {\bibinfo {volume} {11}},\ \bibinfo {pages} {076} (\bibinfo {year} {2021})}\BibitemShut {NoStop}%
\bibitem [{\citenamefont {Gra{\ss}}\ and\ \citenamefont {Lewenstein}(2014)}]{grass2014trapped}%
  \BibitemOpen
  \bibfield  {author} {\bibinfo {author} {\bibfnamefont {T.}~\bibnamefont {Gra{\ss}}}\ and\ \bibinfo {author} {\bibfnamefont {M.}~\bibnamefont {Lewenstein}},\ }\bibfield  {title} {\bibinfo {title} {Trapped-ion quantum simulation of tunable-range heisenberg chains},\ }\href@noop {} {\bibfield  {journal} {\bibinfo  {journal} {EPJ Quantum Technology}\ }\textbf {\bibinfo {volume} {1}},\ \bibinfo {pages} {1} (\bibinfo {year} {2014})}\BibitemShut {NoStop}%
\bibitem [{\citenamefont {Stute}\ \emph {et~al.}(2012)\citenamefont {Stute}, \citenamefont {Casabone}, \citenamefont {Schindler}, \citenamefont {Monz}, \citenamefont {Schmidt}, \citenamefont {Brandst{\"a}tter}, \citenamefont {Northup},\ and\ \citenamefont {Blatt}}]{stute2012tunable}%
  \BibitemOpen
  \bibfield  {author} {\bibinfo {author} {\bibfnamefont {A.}~\bibnamefont {Stute}}, \bibinfo {author} {\bibfnamefont {B.}~\bibnamefont {Casabone}}, \bibinfo {author} {\bibfnamefont {P.}~\bibnamefont {Schindler}}, \bibinfo {author} {\bibfnamefont {T.}~\bibnamefont {Monz}}, \bibinfo {author} {\bibfnamefont {P.~O.}\ \bibnamefont {Schmidt}}, \bibinfo {author} {\bibfnamefont {B.}~\bibnamefont {Brandst{\"a}tter}}, \bibinfo {author} {\bibfnamefont {T.~E.}\ \bibnamefont {Northup}},\ and\ \bibinfo {author} {\bibfnamefont {R.}~\bibnamefont {Blatt}},\ }\bibfield  {title} {\bibinfo {title} {Tunable ion--photon entanglement in an optical cavity},\ }\href@noop {} {\bibfield  {journal} {\bibinfo  {journal} {Nature}\ }\textbf {\bibinfo {volume} {485}},\ \bibinfo {pages} {482} (\bibinfo {year} {2012})}\BibitemShut {NoStop}%
\bibitem [{\citenamefont {Secker}\ \emph {et~al.}(2016)\citenamefont {Secker}, \citenamefont {Gerritsma}, \citenamefont {Glaetzle},\ and\ \citenamefont {Negretti}}]{secker2016controlled}%
  \BibitemOpen
  \bibfield  {author} {\bibinfo {author} {\bibfnamefont {T.}~\bibnamefont {Secker}}, \bibinfo {author} {\bibfnamefont {R.}~\bibnamefont {Gerritsma}}, \bibinfo {author} {\bibfnamefont {A.~W.}\ \bibnamefont {Glaetzle}},\ and\ \bibinfo {author} {\bibfnamefont {A.}~\bibnamefont {Negretti}},\ }\bibfield  {title} {\bibinfo {title} {Controlled long-range interactions between rydberg atoms and ions},\ }\href@noop {} {\bibfield  {journal} {\bibinfo  {journal} {Physical Review A}\ }\textbf {\bibinfo {volume} {94}},\ \bibinfo {pages} {013420} (\bibinfo {year} {2016})}\BibitemShut {NoStop}%
\bibitem [{\citenamefont {Zhang}\ \emph {et~al.}(2017)\citenamefont {Zhang}, \citenamefont {Sun}, \citenamefont {Stowell}, \citenamefont {Zayernouri},\ and\ \citenamefont {Hansen}}]{zhang2017review}%
  \BibitemOpen
  \bibfield  {author} {\bibinfo {author} {\bibfnamefont {Y.}~\bibnamefont {Zhang}}, \bibinfo {author} {\bibfnamefont {H.}~\bibnamefont {Sun}}, \bibinfo {author} {\bibfnamefont {H.~H.}\ \bibnamefont {Stowell}}, \bibinfo {author} {\bibfnamefont {M.}~\bibnamefont {Zayernouri}},\ and\ \bibinfo {author} {\bibfnamefont {S.~E.}\ \bibnamefont {Hansen}},\ }\bibfield  {title} {\bibinfo {title} {A review of applications of fractional calculus in earth system dynamics},\ }\href@noop {} {\bibfield  {journal} {\bibinfo  {journal} {Chaos, Solitons \& Fractals}\ }\textbf {\bibinfo {volume} {102}},\ \bibinfo {pages} {29} (\bibinfo {year} {2017})}\BibitemShut {NoStop}%
\bibitem [{\citenamefont {Chen}(2008)}]{chen2008intuitive}%
  \BibitemOpen
  \bibfield  {author} {\bibinfo {author} {\bibfnamefont {W.}~\bibnamefont {Chen}},\ }\bibfield  {title} {\bibinfo {title} {An intuitive study of fractional derivative modeling and fractional quantum in soft matter},\ }\href@noop {} {\bibfield  {journal} {\bibinfo  {journal} {Journal of Vibration and Control}\ }\textbf {\bibinfo {volume} {14}},\ \bibinfo {pages} {1651} (\bibinfo {year} {2008})}\BibitemShut {NoStop}%
\bibitem [{\citenamefont {Nonnenmacher}\ and\ \citenamefont {Metzler}(2000)}]{nonnenmacher2000applications}%
  \BibitemOpen
  \bibfield  {author} {\bibinfo {author} {\bibfnamefont {T.~F.}\ \bibnamefont {Nonnenmacher}}\ and\ \bibinfo {author} {\bibfnamefont {R.}~\bibnamefont {Metzler}},\ }\bibfield  {title} {\bibinfo {title} {Applications of fractional calculus techniques to problems in biophysics},\ }\href@noop {} {\bibfield  {journal} {\bibinfo  {journal} {Applications of fractional calculus in physics}\ ,\ \bibinfo {pages} {377}} (\bibinfo {year} {2000})}\BibitemShut {NoStop}%
\bibitem [{\citenamefont {Laskin}(2000{\natexlab{a}})}]{laskin2000fractional}%
  \BibitemOpen
  \bibfield  {author} {\bibinfo {author} {\bibfnamefont {N.}~\bibnamefont {Laskin}},\ }\bibfield  {title} {\bibinfo {title} {Fractional quantum mechanics and l{\'e}vy path integrals},\ }\href@noop {} {\bibfield  {journal} {\bibinfo  {journal} {Physics Letters A}\ }\textbf {\bibinfo {volume} {268}},\ \bibinfo {pages} {298} (\bibinfo {year} {2000}{\natexlab{a}})}\BibitemShut {NoStop}%
\bibitem [{\citenamefont {Laskin}(2000{\natexlab{b}})}]{laskin2000fractionalQ}%
  \BibitemOpen
  \bibfield  {author} {\bibinfo {author} {\bibfnamefont {N.}~\bibnamefont {Laskin}},\ }\bibfield  {title} {\bibinfo {title} {Fractional quantum mechanics},\ }\href@noop {} {\bibfield  {journal} {\bibinfo  {journal} {Physical Review E}\ }\textbf {\bibinfo {volume} {62}},\ \bibinfo {pages} {3135} (\bibinfo {year} {2000}{\natexlab{b}})}\BibitemShut {NoStop}%
\bibitem [{\citenamefont {Laskin}(2002)}]{laskin2002fractional}%
  \BibitemOpen
  \bibfield  {author} {\bibinfo {author} {\bibfnamefont {N.}~\bibnamefont {Laskin}},\ }\bibfield  {title} {\bibinfo {title} {Fractional schr{\"o}dinger equation},\ }\href@noop {} {\bibfield  {journal} {\bibinfo  {journal} {Physical Review E}\ }\textbf {\bibinfo {volume} {66}},\ \bibinfo {pages} {056108} (\bibinfo {year} {2002})}\BibitemShut {NoStop}%
\bibitem [{\citenamefont {Stickler}(2013)}]{stickler2013potential}%
  \BibitemOpen
  \bibfield  {author} {\bibinfo {author} {\bibfnamefont {B.}~\bibnamefont {Stickler}},\ }\bibfield  {title} {\bibinfo {title} {Potential condensed-matter realization of space-fractional quantum mechanics: The one-dimensional l{\'e}vy crystal},\ }\href@noop {} {\bibfield  {journal} {\bibinfo  {journal} {Physical Review E}\ }\textbf {\bibinfo {volume} {88}},\ \bibinfo {pages} {012120} (\bibinfo {year} {2013})}\BibitemShut {NoStop}%
\bibitem [{\citenamefont {Hasan}\ and\ \citenamefont {Mandal}(2018)}]{hasan2018tunneling}%
  \BibitemOpen
  \bibfield  {author} {\bibinfo {author} {\bibfnamefont {M.}~\bibnamefont {Hasan}}\ and\ \bibinfo {author} {\bibfnamefont {B.~P.}\ \bibnamefont {Mandal}},\ }\bibfield  {title} {\bibinfo {title} {Tunneling time in space fractional quantum mechanics},\ }\href@noop {} {\bibfield  {journal} {\bibinfo  {journal} {Physics Letters A}\ }\textbf {\bibinfo {volume} {382}},\ \bibinfo {pages} {248} (\bibinfo {year} {2018})}\BibitemShut {NoStop}%
\bibitem [{\citenamefont {Zhang}\ \emph {et~al.}(2015)\citenamefont {Zhang}, \citenamefont {Liu}, \citenamefont {Beli{\'c}}, \citenamefont {Zhong}, \citenamefont {Zhang}, \citenamefont {Xiao} \emph {et~al.}}]{zhang2015propagation}%
  \BibitemOpen
  \bibfield  {author} {\bibinfo {author} {\bibfnamefont {Y.}~\bibnamefont {Zhang}}, \bibinfo {author} {\bibfnamefont {X.}~\bibnamefont {Liu}}, \bibinfo {author} {\bibfnamefont {M.~R.}\ \bibnamefont {Beli{\'c}}}, \bibinfo {author} {\bibfnamefont {W.}~\bibnamefont {Zhong}}, \bibinfo {author} {\bibfnamefont {Y.}~\bibnamefont {Zhang}}, \bibinfo {author} {\bibfnamefont {M.}~\bibnamefont {Xiao}}, \emph {et~al.},\ }\bibfield  {title} {\bibinfo {title} {Propagation dynamics of a light beam in a fractional schr{\"o}dinger equation},\ }\href@noop {} {\bibfield  {journal} {\bibinfo  {journal} {Physical review letters}\ }\textbf {\bibinfo {volume} {115}},\ \bibinfo {pages} {180403} (\bibinfo {year} {2015})}\BibitemShut {NoStop}%
\bibitem [{\citenamefont {Mendl}\ and\ \citenamefont {Spohn}(2015)}]{mendl2015current}%
  \BibitemOpen
  \bibfield  {author} {\bibinfo {author} {\bibfnamefont {C.~B.}\ \bibnamefont {Mendl}}\ and\ \bibinfo {author} {\bibfnamefont {H.}~\bibnamefont {Spohn}},\ }\bibfield  {title} {\bibinfo {title} {Current fluctuations for anharmonic chains in thermal equilibrium},\ }\href@noop {} {\bibfield  {journal} {\bibinfo  {journal} {Journal of Statistical Mechanics: Theory and Experiment}\ }\textbf {\bibinfo {volume} {2015}},\ \bibinfo {pages} {P03007} (\bibinfo {year} {2015})}\BibitemShut {NoStop}%
\bibitem [{\citenamefont {Van~Beijeren}(2012)}]{van2012exact}%
  \BibitemOpen
  \bibfield  {author} {\bibinfo {author} {\bibfnamefont {H.}~\bibnamefont {Van~Beijeren}},\ }\bibfield  {title} {\bibinfo {title} {Exact results for anomalous transport in one-dimensional hamiltonian systems},\ }\href@noop {} {\bibfield  {journal} {\bibinfo  {journal} {Physical review letters}\ }\textbf {\bibinfo {volume} {108}},\ \bibinfo {pages} {180601} (\bibinfo {year} {2012})}\BibitemShut {NoStop}%
\bibitem [{\citenamefont {Dhar}\ \emph {et~al.}(2013)\citenamefont {Dhar}, \citenamefont {Saito},\ and\ \citenamefont {Derrida}}]{dhar2013exact}%
  \BibitemOpen
  \bibfield  {author} {\bibinfo {author} {\bibfnamefont {A.}~\bibnamefont {Dhar}}, \bibinfo {author} {\bibfnamefont {K.}~\bibnamefont {Saito}},\ and\ \bibinfo {author} {\bibfnamefont {B.}~\bibnamefont {Derrida}},\ }\bibfield  {title} {\bibinfo {title} {Exact solution of a l{\'e}vy walk model for anomalous heat transport},\ }\href@noop {} {\bibfield  {journal} {\bibinfo  {journal} {Physical Review E}\ }\textbf {\bibinfo {volume} {87}},\ \bibinfo {pages} {010103} (\bibinfo {year} {2013})}\BibitemShut {NoStop}%
\bibitem [{\citenamefont {Kundu}\ \emph {et~al.}(2019)\citenamefont {Kundu}, \citenamefont {Bernardin}, \citenamefont {Saito}, \citenamefont {Kundu},\ and\ \citenamefont {Dhar}}]{kundu2019fractional}%
  \BibitemOpen
  \bibfield  {author} {\bibinfo {author} {\bibfnamefont {A.}~\bibnamefont {Kundu}}, \bibinfo {author} {\bibfnamefont {C.}~\bibnamefont {Bernardin}}, \bibinfo {author} {\bibfnamefont {K.}~\bibnamefont {Saito}}, \bibinfo {author} {\bibfnamefont {A.}~\bibnamefont {Kundu}},\ and\ \bibinfo {author} {\bibfnamefont {A.}~\bibnamefont {Dhar}},\ }\bibfield  {title} {\bibinfo {title} {Fractional equation description of an open anomalous heat conduction set-up},\ }\href@noop {} {\bibfield  {journal} {\bibinfo  {journal} {Journal of Statistical Mechanics: Theory and Experiment}\ }\textbf {\bibinfo {volume} {2019}},\ \bibinfo {pages} {013205} (\bibinfo {year} {2019})}\BibitemShut {NoStop}%
\bibitem [{\citenamefont {Solomon}\ \emph {et~al.}(1993)\citenamefont {Solomon}, \citenamefont {Weeks},\ and\ \citenamefont {Swinney}}]{solomon1993observation}%
  \BibitemOpen
  \bibfield  {author} {\bibinfo {author} {\bibfnamefont {T.}~\bibnamefont {Solomon}}, \bibinfo {author} {\bibfnamefont {E.~R.}\ \bibnamefont {Weeks}},\ and\ \bibinfo {author} {\bibfnamefont {H.~L.}\ \bibnamefont {Swinney}},\ }\bibfield  {title} {\bibinfo {title} {Observation of anomalous diffusion and l{\'e}vy flights in a two-dimensional rotating flow},\ }\href@noop {} {\bibfield  {journal} {\bibinfo  {journal} {Physical Review Letters}\ }\textbf {\bibinfo {volume} {71}},\ \bibinfo {pages} {3975} (\bibinfo {year} {1993})}\BibitemShut {NoStop}%
\bibitem [{\citenamefont {Pandey}\ and\ \citenamefont {Holm}(2016)}]{pandey2016linking}%
  \BibitemOpen
  \bibfield  {author} {\bibinfo {author} {\bibfnamefont {V.}~\bibnamefont {Pandey}}\ and\ \bibinfo {author} {\bibfnamefont {S.}~\bibnamefont {Holm}},\ }\bibfield  {title} {\bibinfo {title} {Linking the fractional derivative and the lomnitz creep law to non-newtonian time-varying viscosity},\ }\href@noop {} {\bibfield  {journal} {\bibinfo  {journal} {Physical Review E}\ }\textbf {\bibinfo {volume} {94}},\ \bibinfo {pages} {032606} (\bibinfo {year} {2016})}\BibitemShut {NoStop}%
\bibitem [{\citenamefont {Brockmann}\ \emph {et~al.}(2006)\citenamefont {Brockmann}, \citenamefont {Hufnagel},\ and\ \citenamefont {Geisel}}]{brockmann2006scaling}%
  \BibitemOpen
  \bibfield  {author} {\bibinfo {author} {\bibfnamefont {D.}~\bibnamefont {Brockmann}}, \bibinfo {author} {\bibfnamefont {L.}~\bibnamefont {Hufnagel}},\ and\ \bibinfo {author} {\bibfnamefont {T.}~\bibnamefont {Geisel}},\ }\bibfield  {title} {\bibinfo {title} {The scaling laws of human travel},\ }\href@noop {} {\bibfield  {journal} {\bibinfo  {journal} {Nature}\ }\textbf {\bibinfo {volume} {439}},\ \bibinfo {pages} {462} (\bibinfo {year} {2006})}\BibitemShut {NoStop}%
\bibitem [{\citenamefont {Benhamou}(2007)}]{benhamou2007many}%
  \BibitemOpen
  \bibfield  {author} {\bibinfo {author} {\bibfnamefont {S.}~\bibnamefont {Benhamou}},\ }\bibfield  {title} {\bibinfo {title} {How many animals really do the l{\'e}vy walk?},\ }\href@noop {} {\bibfield  {journal} {\bibinfo  {journal} {Ecology}\ }\textbf {\bibinfo {volume} {88}},\ \bibinfo {pages} {1962} (\bibinfo {year} {2007})}\BibitemShut {NoStop}%
\bibitem [{\citenamefont {Murakami}\ \emph {et~al.}(2019)\citenamefont {Murakami}, \citenamefont {Feliciani},\ and\ \citenamefont {Nishinari}}]{murakami2019levy}%
  \BibitemOpen
  \bibfield  {author} {\bibinfo {author} {\bibfnamefont {H.}~\bibnamefont {Murakami}}, \bibinfo {author} {\bibfnamefont {C.}~\bibnamefont {Feliciani}},\ and\ \bibinfo {author} {\bibfnamefont {K.}~\bibnamefont {Nishinari}},\ }\bibfield  {title} {\bibinfo {title} {L{\'e}vy walk process in self-organization of pedestrian crowds},\ }\href@noop {} {\bibfield  {journal} {\bibinfo  {journal} {Journal of The Royal Society Interface}\ }\textbf {\bibinfo {volume} {16}},\ \bibinfo {pages} {20180939} (\bibinfo {year} {2019})}\BibitemShut {NoStop}%
\bibitem [{\citenamefont {Liu}\ \emph {et~al.}(2021)\citenamefont {Liu}, \citenamefont {Long}, \citenamefont {Martin}, \citenamefont {Solomon},\ and\ \citenamefont {Gong}}]{liu2021levy}%
  \BibitemOpen
  \bibfield  {author} {\bibinfo {author} {\bibfnamefont {Y.}~\bibnamefont {Liu}}, \bibinfo {author} {\bibfnamefont {X.}~\bibnamefont {Long}}, \bibinfo {author} {\bibfnamefont {P.~R.}\ \bibnamefont {Martin}}, \bibinfo {author} {\bibfnamefont {S.~G.}\ \bibnamefont {Solomon}},\ and\ \bibinfo {author} {\bibfnamefont {P.}~\bibnamefont {Gong}},\ }\bibfield  {title} {\bibinfo {title} {L{\'e}vy walk dynamics explain gamma burst patterns in primate cerebral cortex},\ }\href@noop {} {\bibfield  {journal} {\bibinfo  {journal} {Communications Biology}\ }\textbf {\bibinfo {volume} {4}},\ \bibinfo {pages} {739} (\bibinfo {year} {2021})}\BibitemShut {NoStop}%
\bibitem [{\citenamefont {Yarahmadi}\ and\ \citenamefont {Saberi}(2022)}]{yarahmadi20222d}%
  \BibitemOpen
  \bibfield  {author} {\bibinfo {author} {\bibfnamefont {H.}~\bibnamefont {Yarahmadi}}\ and\ \bibinfo {author} {\bibfnamefont {A.~A.}\ \bibnamefont {Saberi}},\ }\bibfield  {title} {\bibinfo {title} {A 2d l{\'e}vy-flight model for the complex dynamics of real-life financial markets},\ }\href@noop {} {\bibfield  {journal} {\bibinfo  {journal} {Chaos: An Interdisciplinary Journal of Nonlinear Science}\ }\textbf {\bibinfo {volume} {32}} (\bibinfo {year} {2022})}\BibitemShut {NoStop}%
\bibitem [{\citenamefont {Iomin}(2021)}]{iomin2021fractional}%
  \BibitemOpen
  \bibfield  {author} {\bibinfo {author} {\bibfnamefont {A.}~\bibnamefont {Iomin}},\ }\bibfield  {title} {\bibinfo {title} {Fractional schr{\"o}dinger equation in gravitational optics},\ }\href@noop {} {\bibfield  {journal} {\bibinfo  {journal} {Modern Physics Letters A}\ }\textbf {\bibinfo {volume} {36}},\ \bibinfo {pages} {2140003} (\bibinfo {year} {2021})}\BibitemShut {NoStop}%
\bibitem [{\citenamefont {Zeng}\ and\ \citenamefont {Zeng}(2019)}]{zeng2019one}%
  \BibitemOpen
  \bibfield  {author} {\bibinfo {author} {\bibfnamefont {L.}~\bibnamefont {Zeng}}\ and\ \bibinfo {author} {\bibfnamefont {J.}~\bibnamefont {Zeng}},\ }\bibfield  {title} {\bibinfo {title} {One-dimensional solitons in fractional schr{\"o}dinger equation with a spatially periodical modulated nonlinearity: nonlinear lattice},\ }\href@noop {} {\bibfield  {journal} {\bibinfo  {journal} {Optics Letters}\ }\textbf {\bibinfo {volume} {44}},\ \bibinfo {pages} {2661} (\bibinfo {year} {2019})}\BibitemShut {NoStop}%
\bibitem [{\citenamefont {Xin}\ \emph {et~al.}(2021)\citenamefont {Xin}, \citenamefont {Song},\ and\ \citenamefont {Li}}]{xin2021propagation}%
  \BibitemOpen
  \bibfield  {author} {\bibinfo {author} {\bibfnamefont {W.}~\bibnamefont {Xin}}, \bibinfo {author} {\bibfnamefont {L.}~\bibnamefont {Song}},\ and\ \bibinfo {author} {\bibfnamefont {L.}~\bibnamefont {Li}},\ }\bibfield  {title} {\bibinfo {title} {Propagation of gaussian beam based on two-dimensional fractional schr{\"o}dinger equation},\ }\href@noop {} {\bibfield  {journal} {\bibinfo  {journal} {Optics Communications}\ }\textbf {\bibinfo {volume} {480}},\ \bibinfo {pages} {126483} (\bibinfo {year} {2021})}\BibitemShut {NoStop}%
\bibitem [{\citenamefont {He}\ \emph {et~al.}(2021)\citenamefont {He}, \citenamefont {Malomed}, \citenamefont {Mihalache}, \citenamefont {Peng}, \citenamefont {He},\ and\ \citenamefont {Deng}}]{he2021propagation}%
  \BibitemOpen
  \bibfield  {author} {\bibinfo {author} {\bibfnamefont {S.}~\bibnamefont {He}}, \bibinfo {author} {\bibfnamefont {B.~A.}\ \bibnamefont {Malomed}}, \bibinfo {author} {\bibfnamefont {D.}~\bibnamefont {Mihalache}}, \bibinfo {author} {\bibfnamefont {X.}~\bibnamefont {Peng}}, \bibinfo {author} {\bibfnamefont {Y.}~\bibnamefont {He}},\ and\ \bibinfo {author} {\bibfnamefont {D.}~\bibnamefont {Deng}},\ }\bibfield  {title} {\bibinfo {title} {Propagation dynamics of radially polarized symmetric airy beams in the fractional schr{\"o}dinger equation},\ }\href@noop {} {\bibfield  {journal} {\bibinfo  {journal} {Physics Letters A}\ }\textbf {\bibinfo {volume} {404}},\ \bibinfo {pages} {127403} (\bibinfo {year} {2021})}\BibitemShut {NoStop}%
\bibitem [{\citenamefont {Liu}\ \emph {et~al.}(2023)\citenamefont {Liu}, \citenamefont {Zhang}, \citenamefont {Malomed},\ and\ \citenamefont {Karimi}}]{liu2023experimental}%
  \BibitemOpen
  \bibfield  {author} {\bibinfo {author} {\bibfnamefont {S.}~\bibnamefont {Liu}}, \bibinfo {author} {\bibfnamefont {Y.}~\bibnamefont {Zhang}}, \bibinfo {author} {\bibfnamefont {B.~A.}\ \bibnamefont {Malomed}},\ and\ \bibinfo {author} {\bibfnamefont {E.}~\bibnamefont {Karimi}},\ }\bibfield  {title} {\bibinfo {title} {Experimental realisations of the fractional schr{\"o}dinger equation in the temporal domain},\ }\href@noop {} {\bibfield  {journal} {\bibinfo  {journal} {Nature Communications}\ }\textbf {\bibinfo {volume} {14}},\ \bibinfo {pages} {222} (\bibinfo {year} {2023})}\BibitemShut {NoStop}%
\bibitem [{\citenamefont {Ortigueira}(2006)}]{ortigueira2006riesz}%
  \BibitemOpen
  \bibfield  {author} {\bibinfo {author} {\bibfnamefont {M.~D.}\ \bibnamefont {Ortigueira}},\ }\bibfield  {title} {\bibinfo {title} {Riesz potential operators and inverses via fractional centred derivatives},\ }\href@noop {} {\bibfield  {journal} {\bibinfo  {journal} {International Journal of Mathematics and Mathematical Sciences}\ }\textbf {\bibinfo {volume} {2006}},\ \bibinfo {pages} {048391} (\bibinfo {year} {2006})}\BibitemShut {NoStop}%
\bibitem [{\citenamefont {Privman}(1990)}]{privman1990finite}%
  \BibitemOpen
  \bibfield  {author} {\bibinfo {author} {\bibfnamefont {V.}~\bibnamefont {Privman}},\ }\href@noop {} {\emph {\bibinfo {title} {Finite size scaling and numerical simulation of statistical systems}}}\ (\bibinfo  {publisher} {World Scientific},\ \bibinfo {year} {1990})\BibitemShut {NoStop}%
\bibitem [{\citenamefont {Flores-Sola}\ \emph {et~al.}(2015)\citenamefont {Flores-Sola}, \citenamefont {Berche}, \citenamefont {Kenna},\ and\ \citenamefont {Weigel}}]{flores2015finite}%
  \BibitemOpen
  \bibfield  {author} {\bibinfo {author} {\bibfnamefont {E.~J.}\ \bibnamefont {Flores-Sola}}, \bibinfo {author} {\bibfnamefont {B.}~\bibnamefont {Berche}}, \bibinfo {author} {\bibfnamefont {R.}~\bibnamefont {Kenna}},\ and\ \bibinfo {author} {\bibfnamefont {M.}~\bibnamefont {Weigel}},\ }\bibfield  {title} {\bibinfo {title} {Finite-size scaling above the upper critical dimension in ising models with long-range interactions},\ }\href@noop {} {\bibfield  {journal} {\bibinfo  {journal} {The European Physical Journal B}\ }\textbf {\bibinfo {volume} {88}},\ \bibinfo {pages} {1} (\bibinfo {year} {2015})}\BibitemShut {NoStop}%
\bibitem [{\citenamefont {Hove}\ \emph {et~al.}(2000)\citenamefont {Hove}, \citenamefont {Mo},\ and\ \citenamefont {Sudb{\o}}}]{hove2000hausdorff}%
  \BibitemOpen
  \bibfield  {author} {\bibinfo {author} {\bibfnamefont {J.}~\bibnamefont {Hove}}, \bibinfo {author} {\bibfnamefont {S.}~\bibnamefont {Mo}},\ and\ \bibinfo {author} {\bibfnamefont {A.}~\bibnamefont {Sudb{\o}}},\ }\bibfield  {title} {\bibinfo {title} {Hausdorff dimension of critical fluctuations in abelian gauge theories},\ }\href@noop {} {\bibfield  {journal} {\bibinfo  {journal} {Physical Review Letters}\ }\textbf {\bibinfo {volume} {85}},\ \bibinfo {pages} {2368} (\bibinfo {year} {2000})}\BibitemShut {NoStop}%
\bibitem [{\citenamefont {Vojta}(2003)}]{vojta2003quantum}%
  \BibitemOpen
  \bibfield  {author} {\bibinfo {author} {\bibfnamefont {M.}~\bibnamefont {Vojta}},\ }\bibfield  {title} {\bibinfo {title} {Quantum phase transitions},\ }\href@noop {} {\bibfield  {journal} {\bibinfo  {journal} {Reports on Progress in Physics}\ }\textbf {\bibinfo {volume} {66}},\ \bibinfo {pages} {2069} (\bibinfo {year} {2003})}\BibitemShut {NoStop}%
\end{thebibliography}%

\end{document}